\documentclass[preprints,article,accept,pdftex,moreauthors]{Definitions/mdpi} 
\usepackage{amsmath}
\usepackage{amssymb}
\usepackage{graphicx}
\usepackage{color}
\usepackage{subfig}
\usepackage{bm}
\usepackage{gensymb}
\usepackage{bigints}
\usepackage{pifont}
\usepackage{aas_macros}
\usepackage{physics}
\firstpage{1} 
\pubvolume{1}
\issuenum{1}
\articlenumber{0}
\pubyear{2022}
\copyrightyear{2022}
\datereceived{} 
\dateaccepted{} 
\datepublished{} 
\hreflink{https://doi.org/} 
\pdfoutput=1

\Title{Massive Neutron Stars and White Dwarfs as Noncommutative Fuzzy Spheres}

\TitleCitation{Massive Neutron Stars and White Dwarfs as Noncommutative Fuzzy Spheres}

\Author{Surajit Kalita$^{1,2}$\orcidA{} and Banibrata Mukhopadhyay$^{2}$\orcidB{}}

\AuthorNames{Surajit Kalita and Banibrata Mukhopadhyay}

\AuthorCitation{Kalita, S. and Mukhopadhyay, B.}

\address{%
$^{1}$ \quad High Energy Physics, Cosmology \& Astrophysics Theory (HEPCAT) Group, Department of Mathematics \& Applied Mathematics, University of Cape Town, Cape Town 7700, South Africa; surajit.kalita@uct.ac.za\\
$^{2}$ \quad Department of Physics, Indian Institute of Science, Bangalore 560012, India; bm@iisc.ac.in}

\corres{Correspondence: bm@iisc.ac.in}

\abstract{Over the last couple of decades, there are direct and indirect evidences for massive compact objects than their conventional counterparts. A couple of such examples are super-Chandrasekhar white dwarfs and massive neutron stars. The observations of more than a dozen peculiar over-luminous type Ia supernovae predict their origins from super-Chandrasekhar white dwarf progenitors. On the other hand, recent gravitational wave detection and some pulsar observations argue for massive neutron stars, lying in the famous mass-gap between lowest astrophysical black hole and conventional highest neutron star masses. We show that the idea of a squashed fuzzy sphere, which brings in noncommutative geometry, can self-consistently explain either of the massive objects as if they are actually fuzzy or squashed fuzzy spheres. Noncommutative geometry is a branch of quantum gravity. If the above proposal is correct, it will provide observational evidences for noncommutativity.}


\begin{document}


\section{Introduction}
Quantum mechanics (QM) and general theory of relativity (GR) are widely regarded as the two most promising discoveries of the twentieth century. QM is used to describe different microscopic phenomena, whereas GR is used to explain phenomena in which gravity plays a significant role. QM is primarily based on the Heisenberg algebra, which relates the position operator ($\hat{x}_i$) and the momentum operator ($\hat{p}_i$) as $\comm{\hat{x}_i}{\hat{p}_j}= i\hbar\delta_{ij}$, where $\hbar = h/2\pi$ with $h$ being the Planck constant. Note that in QM, position and momentum operators commute among themselves, i.e. $\comm{\hat{x}_i}{\hat{x}_j}= \comm{\hat{p}_i}{\hat{p}_j}= 0$. GR, on the other hand, is based on the equivalence principle, which can account for the perihelion precision of Mercury, the generation of gravitational waves (GWs), gravitational lensing, and a variety of other fascinating phenomena. Both QM and GR are required to understand the structure of compact objects, such as white dwarfs (WDs) and neutron stars (NSs). GR primarily governs the hydrostatic balance of a star, which is a macroscopic property; whereas QM determines the equation of state (EoS), i.e. the relation between pressure and density of the constituent particles.

If a progenitor star has mass approximately in between $10$ and $20\rm\,M_\odot$, it becomes a NS at the end of its lifetime. A NS typically possesses central density, $\rho_\mathrm{c}$ of about $10^{14}$ to a few factors of $10^{15}\rm\,g\,cm^{-3}$~\cite{compact}. Although NSs predominantly consist of neutrons, various other particles, including hyperons, may also be present at such a high density. This uncertainty arises from the fact that such a high density has yet to be achieved in the laboratory, and hence the specific nuclear reactions, as well as their rates, are unknown. Researchers have so far provided various NS EoSs, each comprising different particle contributions and strong nuclear forces. Most of these EoSs are based on the relativistic energy dispersion relation $E^2 = p^2c^2 + m^2c^4$, where $c$ is the speed of light and $E$ denotes the energy of the particle with mass $m$ with $p$ being its momentum. Although most NSs have masses of approximately $1$ to $2\rm\,M_\odot$, recent pulsar observations PSR\,J2215+5135 and PSR\,B1957+20 show that they have masses of about $2.3$ and $2.4\rm\,M_\odot$, respectively~\cite{2018ApJ...859...54L,2011ApJ...728...95V}. Similarly, the LIGO/Virgo collaboration detected a GW merger event, GW\,190814, where one of the merged objects has a mass of about $2.6\rm\,M_\odot$~\cite{2020ApJ...896L..44A}, which is mostly thought to be a NS~\cite{2020MNRAS.499L..82M,2020ApJ...904...39H,2020ApJ...905...48T,2021PhRvC.103b5808D}. Nevertheless, there was no detection of electromagnetic counterpart for this GW event, and hence various other proposals for this object, such as black hole~\cite{2020ApJ...901L..34Y,2020PhRvD.102f1301V}, quark star~\cite{2021PhRvL.126p2702B}, etc., have been put forward. In this article, however, we only talk about NSs while referring to this GW event. Based on these observations, various simulations have been performed and it has been suggested that those EoSs, which give the maximum mass of a non-rotating and non-magnetized NS less than $2\rm\,M_\odot$, should be ruled out~\cite{2018ApJ...852L..25R,2019PhRvC..99e2801M,2020MNRAS.499L..82M}. Hence, considering GR formalism, various EoSs, such as FPS~\cite{1989ASIB..205..103P}, ALF1~\cite{2005ApJ...629..969A}, etc., seem to be inappropriate for NSs. Modified gravity, on the other hand, has emerged as a popular alternative to replace GR in the high-density regime over the last decades. It can be shown that modified gravity alters the hydrostatic balance of the star and thereby increases the mass of a NS~\cite{2014PhRvD..89j3509A,2017CQGra..34t5008A,2016IJMPS..4160130A}. As a result, some of these EoSs may still be valid in the modified gravity formalism.

On the other hand, WDs are the end-state of stars with mass  $\lesssim(10\pm2)\rm\,M_\odot$~\cite{2018MNRAS.480.1547L}. They possess $\rho_\mathrm{c}$ typically ranging approximately from $10^{5}\rm\,g\,cm^{-3}$ to a few factor of $10^{10}\rm\,g\,cm^{-3}$. A WD achieves its stable equilibrium configuration by balancing the outward force of the degenerate electron gas with the inward force of gravity. If the WD has a binary companion, it pulls out matter from the companion, resulting in the increase of WD mass. Once the WD hits the Chandrasekhar mass-limit, which is about $1.4\rm\,M_\odot$ for a carbon–oxygen non-rotating, nonmagnetized WD~\cite{1935MNRAS..95..207C}, this pressure balance is lost, and it bursts out to create a type Ia supernova (SN\,Ia). However, recent observations of more than a dozen of peculiar over-luminous SNe\,Ia~\cite{2006Natur.443..308H,2007ApJ...669L..17H,2009ApJ...707L.118Y,2010ApJ...714.1209T,2011MNRAS.410..585S,2011MNRAS.412.2735T,2010ApJ...715.1338Y,2012ApJ...757...12S,2016ApJ...823..147C} reveal that they had to be produced from super-Chandrasekhar limiting mass WDs, i.e. the WDs burst significantly above the Chandrasekhar mass-limit~\cite{2010ApJ...713.1073S,2012ApJ...756..191K}. Various theories incorporating magnetic fields~\cite{2013PhRvL.110g1102D,2019MNRAS.490.2692K}, modified gravity~\cite{2018JCAP...09..007K,2022PhRvD.105b4028S,2022PhLB..82736942K}, etc. can explain this violation of the Chandrasekhar mass-limit, albeit each has its own set of limitations.

The goal of this work is to introduce noncommutativity (NC) among position and momentum variables and examine how it affects WDs and NSs. A popular way of proposing NC is by defining $\comm{x_i}{x_j}= i\eta$ and $\comm{p_i}{p_j}= i\theta$ with $\eta$ and $\theta$ being the NC parameters. It was shown that in the presence of NC, the spacetime metric alters~\cite{2006LNP...698...97N}; causing the event horizon to shift and the singularity at the centre of a black hole to vanish, which is replaced by a regular de-Sitter core~\cite{2006PhLB..632..547N,2009IJMPA..24.1229N,2017EPJC...77..577K}. It further alters some other properties associated with black holes, such as the stability of Cauchy horizon~\cite{2010PhLB..692...32B}, mini black hole formation with the central singularity replaced by a self-gravitating droplet~\cite{2009CQGra..26x5006A}, the Hawking temperature~\cite{2018EPJP..133..421F}. Various researchers also utilised this NC to describe a variety of other phenomena, including Berry curvature, fundamental length-scale, Landau levels, gamma-ray bursts, and many more~\cite{1999JHEP...09..032S,2000IJMPA..15.4301A,2001PhLB..510..255A,2002PhRvL..88s0403M,2002Natur.418...34A,2012PhRvL.109r1602S}. Note that the basic assumption in the structure of this NC is quite ad-hoc. In 1992, Madore introduced the idea of a 3-dimensional fuzzy sphere NC~\cite{1992CQGra...9...69M}, which has been used to better understand the thermodynamical features of non-interacting degenerate electron gas~\cite{2014JPhA...47R5203C,2015PhRvD..92l5013S}. This formalism was later refined by projecting all the points of the fuzzy sphere onto an equatorial plane and named this configuration a squashed fuzzy sphere~\cite{2015JPhA...48C5401A}. This NC model was also proven to imitate the magnetic field by producing distinct energy levels, which are similar to the Landau levels created in the presence of a magnetic field~\cite{1991ApJ...383..745L}.

Apart from a few black hole applications, the implication of NC on compact objects is a relatively novel concept. We earlier showed its applications on the structure of WDs. We considered both the formalism of NC separately and showed that they modify the energy dispersion relation of electrons~\cite{2021IJMPD..3050034K,2021IJMPD..3050101K}. We further used this relation to obtain a new EoS of the degenerate electrons present in WDs and showed that it can explain the super-Chandrasekhar limiting mass WDs, which are believed to be the progenitors of the observed over-luminous type Ia supernovae. We obtained the maximum mass of a WD to be about $2.6\rm\,M_\odot$ in the presence of NC, and this mass-limit decreases as the strength of NC reduces. We further showed that the NC is prominent if the separation of electrons is less than the Compton wavelength of electrons, and it turns out to be an emergent phenomenon.

The EoS obtained for WD is valid only up to neutron drip density, above which neutron starts contributing to the degenerate pressure. In this article, we obtain a new EoS above the neutron drip density taking into account of NC and derive a new mass--radius relation for NSs. With the advancement of technology, different proposed electromagnetic and GW detectors are likely to detect numerous WDs and NSs. If their observed masses and radii follow the mass--radius relations predicted based on NC, it would be a direct proof of NC's existence. 

The following is a breakdown of how this article is structured. In Section~\ref{Sec: 2}, we briefly review the squashed fuzzy sphere formalism and the modified energy dispersion relation, which we utilize in Section~\ref{Sec: 3} to derive the EoS for degenerate particles reside inside WD and NS in the presence of NC. We further use this EoS to obtain the new mass--radius relation of the NS in Section~\ref{Sec: 4}. Finally, we put our concluding remarks in Section~\ref{Sec: 5}.

\section{Squashed fuzzy sphere formalism and modified energy dispersion relation}\label{Sec: 2}

In this section, we recapitulate the basic formalism of a squashed fuzzy sphere. In $\mathbb{R}^3$, the equation of a sphere with radius $r$ is given by
\begin{equation}\label{Eq: sphere}
x_1^2 + x_2^2 + x_3^2 = r^2,
\end{equation}
where $\left(x_1, x_2, x_3\right)$ are the Cartesian coordinates of the points on the sphere. A fuzzy sphere is similar to a regular sphere, except that its coordinates $x_i$ ($i=1,2,3$) follow the regular QM angular momentum algebra~\cite{1992CQGra...9...69M}. Hence, if $J_i$ are the generators of $\mathtt{SU}(2)$ group in an $N$-dimensional irreducible representation, we have
\begin{equation}
x_i = \kappa J_i,
\end{equation}
with
\begin{equation}
J_1^2 + J_2^2 + J_3^2 = \frac{\hbar^2}{4}\left(N^2-1\right) \mathbb{I} = C_N \mathbb{I},
\end{equation}
where $\kappa$ is the proportionality (scaling) constant, $C_N = \hbar^2\left(N^2-1\right)/4$, and $\mathbb{I}$ is the $N$-dimensional identity matrix. Substituting $J_i$ in terms of $x_i$ and defining $k = \kappa r$, we obtain
\begin{align}\label{Eq: k_R relation}
	\kappa = \frac{r}{\sqrt{C_N}} \quad \mathrm{and} \quad k = \frac{r^2}{\sqrt{C_N}}.
\end{align}
Since the angular momentum algebra follows the commutation relation $\comm{J_j}{J_k} = i\hbar\epsilon_{jkl}J_l$, the coordinates of the fuzzy sphere follow \cite{1992CQGra...9...69M}
\begin{equation}\label{Eq: Fuzzy sphere}
\comm{x_j}{x_k} = i\frac{k \hbar}{r} \epsilon_{jkl} x_l.
\end{equation}

\begin{figure}[htpb]
	\centering
	\includegraphics[scale=0.5]{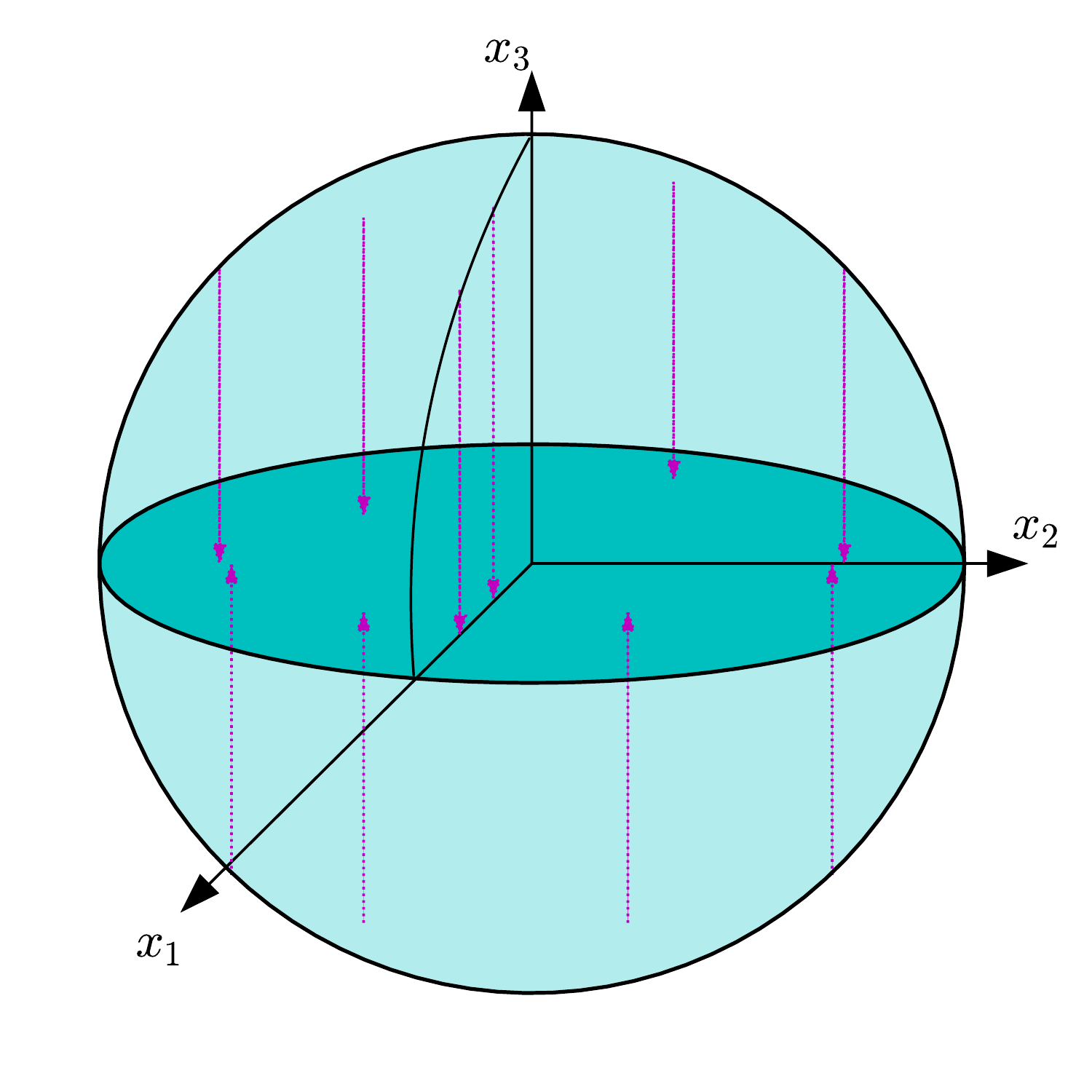}
	\caption{Schematic diagram of a squashed fuzzy sphere where all the points of the fuzzy sphere are projected on $x_1$-$x_2$ plane.}
	\label{Fig: squashed fuzzy sphere}
\end{figure}
When all the points of a fuzzy sphere are projected on any of its equatorial planes, the result is a squashed fuzzy sphere. It should be noted that this is not a stereographic projection. The projection of all the points of a fuzzy sphere on the $x_1$-$x_2$ equatorial plane is shown in Figure~\ref{Fig: squashed fuzzy sphere}. The points of the upper hemisphere are projected on the equatorial plane's top side, while the points of the lower hemisphere are projected on the plane's lower side, and then they are glued together. Writing $x_3$ in terms of $x_1$ and $x_2$ using Equation~\eqref{Eq: sphere} and replacing it in Equation~\eqref{Eq: Fuzzy sphere}, we obtain the squashed fuzzy sphere's commutation relation, given by~\cite{2015JPhA...48C5401A}
\begin{equation}\label{Eq: Squashed fuzzy sphere}
\comm{x_1}{x_2} = \pm i \frac{k \hbar}{r} \sqrt{r^2 - x_1^2 - x_2^2}.
\end{equation}
The Laplacian for the squashed fuzzy sphere is given by \cite{2015JPhA...48C5401A}
\begin{equation}
\Box_s = \frac{1}{k^2}\sum_{i=1}^2 \comm{X_i}{\comm{X_i}{\cdot}},
\end{equation}
which satisfies the following eigenvalue equation
\begin{equation}
\Box_s \hat{Y}^{\Tilde{l}}_{\Tilde{m}} = \frac{\hbar^2}{r^2}\left\{\tilde{l}(\tilde{l}+1)-\tilde{m}^2 \right\}\hat{Y}^{\Tilde{l}}_{\Tilde{m}},
\end{equation}
where $\tilde{l}(\tilde{l}+1)-\tilde{m}^2$ are eigenvalues of the squashed fuzzy Laplacian with $\tilde{l}$ taking all the integer values from $0$ to $N-1$ and $\tilde{m}$ taking all the integer values from $-\tilde{l}$ to $\tilde{l}$. Using this Laplacian, one can obtain the energy dispersion relation in the squashed fuzzy sphere, given by \cite{2015JPhA...48C5401A,2021IJMPD..3050101K}
\begin{align} \label{Eq: enrgy eigenvalues}
	E_{\tilde{l},\tilde{m}}^2 = \frac{2 \hbar c^2}{k \sqrt{N^2-1}} \left[\tilde{l}(\tilde{l}+1)-\tilde{m}(\tilde{m}\pm 1)\right].
\end{align}
Moreover, Equation~\eqref{Eq: Squashed fuzzy sphere} in spherical polar coordinates $(r,\theta,\phi)$ can be recast as
\begin{align}
	\comm{\sin\theta \cos\phi}{\sin\theta \sin\phi} &= \pm i \frac{k \hbar}{r^2} \cos\theta.
\end{align}
This shows NC is between $\theta$ and $\phi$ alone, while they are commutative with $r$-coordinate. In other words, the formalism of squashed fuzzy sphere is such that it actually provides a NC between the azimuthal and polar coordinates. This is because the squashed plane in a fuzzy sphere can be any of its equatorial planes, which means that the squashed fuzzy sphere possesses rotational symmetry about the equatorial plane. Regardless of the squashed plane, the above energy dispersion remains unchanged. As a result, a particle traveling along the $r$-coordinate in a squashed fuzzy sphere is not affected by NC and the exact energy dispersion relation is given by
\begin{align}\label{Eq: Fuzzy dispersion relation: v1}
	E^2 = p_r^2 c^2 + m^2 c^4 \left[1+ \{\tilde{l}(\tilde{l}+1)-\tilde{m}(\tilde{m} \pm 1)\} \frac{2 \hbar}{m^2 c^2 k \sqrt{N^2-1}}\right],
\end{align}
where $p_r$ is the momentum of the particle in the radial direction. In the limit $N\gg1$, the above expression reduces to \cite{2021IJMPD..3050101K}
\begin{align}\label{Eq: Fuzzy dispersion relation}
	E^2 = p_r^2 c^2 + m^2 c^4 \left(1+ 2\nu \theta_\mathrm{D}\right),\qquad \nu \in \mathbb{Z}^{0+},
\end{align}
where $\theta_\mathrm{D} = 2 \hbar/m^2 c^2 k$.
It is noticeable that this expression is very similar to the dispersion relation of Landau levels in the presence of a magnetic field. If the magnetic field is present along $z$-direction with strength $B$, the energy dispersion relation for an electron with mass $m_\mathrm{e}$ is given by~\cite{1991ApJ...383..745L}
\begin{align}\label{Eq: Magnetic EoS}
	E^2 &= p_z^2c^2 + m_\mathrm{e}^2 c^4 \left(1+ 2\nu \frac{B}{B_\mathrm{c}}\right),\qquad \nu \in \mathbb{Z}^{0+},
\end{align}
where $p_z$ is the momentum of the electron along the $z$-direction and $B_\mathrm{c} = m_\mathrm{e}^2 c^3/\hbar e$ is the critical magnetic field (Schwinger limit) with $e$ being the charge of an electron. Comparing Equations~\eqref{Eq: Fuzzy dispersion relation} and \eqref{Eq: Magnetic EoS}, we obtain
\begin{align}
	B \equiv \frac{2c}{ek}.
\end{align}
Hence, in a squashed fuzzy sphere, $k^{-1}$ behaves as the strength of NC. A detailed discussion on the equivalence of magnetic field and NC was given by Kalita et al.~\cite{2021IJMPD..3050101K}. Equation~\eqref{Eq: Fuzzy dispersion relation} provides the energy dispersion relation of one squashed fuzzy sphere, inside which $k$ is constant. If we consider a sequence of concentric squashed fuzzy spheres with same $N$, from Equation~\eqref{Eq: k_R relation}, we have $k\propto r^2$, i.e. $k$ increases and thus the strength of NC reduces from center to the surface. As a result, all concentric spheres with a radius greater than $r$ contribute to the effective NC at a point with radius $r$. From Equation~\eqref{Eq: Squashed fuzzy sphere}, it is evident that NC vanishes at the surface.

\section{Noncommutative equation of state for degenerate particles}\label{Sec: 3}

In this section, we first discuss the commutative cases.
In 1935, Chandrasekhar provided EoS for the degenerate electrons \cite{1935MNRAS..95..207C}. This EoS is valid for a system whose density is less than the neutron drip density (approximately $3.18\times10^{11}\rm\,g\,cm^{-3}$), above which neutron also starts contributing to the degenerate pressure. Harrison and Wheeler (hereinafter HW), in 1958, provided an EoS considering a semi-empirical mass formula, which is valid even at higher densities than neutron drip density. Denoting $\rho$ to be the matter density and $\mathcal{P}$ the total pressure, HW EoS is given by~\cite{compact}
\begin{equation}\label{Eq: HW}
\begin{aligned}
	\rho &= \frac{n_\mathrm{ion} M(A,Z) + \epsilon_\mathrm{e}(n_\mathrm{e}) - n_\mathrm{e}m_\mathrm{e}c^2 + \epsilon_\mathrm{n}(n_\mathrm{n})}{c^2},\\
	\mathcal{P} &= \mathcal{P}_\mathrm{e} + \mathcal{P}_\mathrm{n},
\end{aligned}
\end{equation}
where $\epsilon_\mathrm{n}$ is the energy density of neutrons and $\epsilon_\mathrm{e}$ is the same for electrons. Similarly, $\mathcal{P}_\mathrm{e}$ and $\mathcal{P}_\mathrm{n}$ are respectively the pressures  due to electrons and neutrons. Here $n_\mathrm{e}$, $n_\mathrm{n}$, and $n_\mathrm{ion}$ are the number densities of electron, neutron, and ion respectively, while $M(A,Z)$ is the energy of nucleus with mass number $A$ and atomic number $Z$.

In commutative physics where $E^2 = p^2c^2 + m^2c^4$ holds good, the pressures and energy densities are given by
\begin{align}
	\mathcal{P}_\mathrm{e} &= \frac{m_\mathrm{e}c^2}{\lambda_\mathrm{e}^3}\phi(x_\mathrm{F,e}), \quad \mathcal{P}_\mathrm{n} = \frac{m_\mathrm{n}c^2}{\lambda_\mathrm{n}^3}\phi(x_\mathrm{F,n}),\quad \mathcal{P}_\mathrm{p} = \frac{m_\mathrm{p}c^2}{\lambda_\mathrm{p}^3}\phi(x_\mathrm{F,p}),\\
	\epsilon_\mathrm{e} &= \frac{m_\mathrm{e}c^2}{\lambda_\mathrm{e}^3}\chi(x_\mathrm{F,e}), \quad \epsilon_\mathrm{n} = \frac{m_\mathrm{n}c^2}{\lambda_\mathrm{n}^3}\chi(x_\mathrm{F,n}), \quad \epsilon_\mathrm{p} = \frac{m_\mathrm{p}c^2}{\lambda_\mathrm{p}^3}\chi(x_\mathrm{F,p}),
\end{align}
where $\lambda_\mathrm{e} = \hbar/m_\mathrm{e}c$, $\lambda_\mathrm{n} = \hbar/m_\mathrm{n}c$, and $\lambda_\mathrm{p} = \hbar/m_\mathrm{p}c$ are the reduced Compton wavelengths of electron, neutron, and proton respectively with $m_\mathrm{n}$ being the mass of a neutron and $m_\mathrm{p}$ the mass of a proton. Moreover, $x_\mathrm{F,e}=p_\mathrm{F,e}/m_\mathrm{e}c$, $x_\mathrm{F,n}=p_\mathrm{F,n}/m_\mathrm{n}c$, and $x_\mathrm{F,p}=p_\mathrm{F,p}/m_\mathrm{p}c$ with $p_\mathrm{F,e}$, $p_\mathrm{F,n}$, and $p_\mathrm{F,p}$ being the Fermi momentum of electron, neutron, and proton respectively, and
\begin{align*}
	\phi(x_\mathrm{F}) &= \frac{1}{8\pi^2}\left[x_\mathrm{F}\sqrt{1+x_\mathrm{F}^2} \left(\frac{2x_\mathrm{F}^2}{3}-1\right) + \ln\left\{x_\mathrm{F}+\sqrt{1+x_\mathrm{F}^2}\right\} \right],\\
	\chi(x_\mathrm{F}) &= \frac{1}{8\pi^2}\left[x_\mathrm{F}\sqrt{1+x_\mathrm{F}^2} \left(2x_\mathrm{F}^2+1\right) - \ln\left\{x_\mathrm{F}+\sqrt{1+x_\mathrm{F}^2}\right\} \right].
\end{align*}
This EoS can explain physics beyond the neutron drip density regime. However, above $4.54\times10^{12}\rm\,g\,cm^{-3}$, the neutrons contribute most in the pressure and density. Hence, beyond this density, HW used the idea n-p-e EoS where neutrons, protons, and electrons are considered to be degenerate and non-interacting. In the commutative picture, the n-p-e EoS is given by~\cite{compact}
\begin{equation}\label{Eq: npe}
\begin{aligned}
	\mathcal{P} &= \mathcal{P}_\mathrm{e} + \mathcal{P}_\mathrm{n} + \mathcal{P}_\mathrm{p},\\
	\rho &= \frac{\epsilon_\mathrm{e} + \epsilon_\mathrm{n} + \epsilon_\mathrm{p}}{c^2}.
\end{aligned}
\end{equation}
HW and n-p-e EoSs together provide the pressure--density relation of the non-interacting degenerate particles.

In NC, these EoSs are expected to be modified. Vishal and Mukhopadhyay earlier derived a modified HW EoS of degenerate particles in the presence of a constant magnetic field~\cite{2014PhRvC..89f5804V}. Later, to study the effect of varying NC on degenerate electron gas, we obtained the following relation~\cite{2021IJMPD..3050101K}
\begin{equation}\label{Eq: theta_D n_e relation}
\theta_\mathrm{D} = \frac{1}{\xi}\frac{h^2n_\mathrm{e}^{2/3}}{\pi m_\mathrm{e}^2c^2},
\end{equation}
where $\xi$ is a dimensionless proportionality constant. The dependency $\theta_\mathrm{D}\propto n_\mathrm{e}^{2/3}$ is required to match the modified EoS with the Chandrasekhar EoS at a low density where NC does not have any significant influence. Thus we obtained the modified EoS for degenerate electrons when all the electrons reside in the ground level, given by~\cite{2021IJMPD..3050034K,2021IJMPD..3050101K}
\begin{align} \label{Eq: Fuzzy pressure}
	\mathcal{P}_\mathrm{e} &= \frac{2 \rho_\mathrm{e}^{2/3}}{\xi h \mu_\mathrm{e}^{2/3} m_\mathrm{p}^{2/3}} \left\{p_\mathrm{F,e} E_\mathrm{F,e} - m_\mathrm{e}^2 c^3 \ln(\frac{E_\mathrm{F,e} + p_\mathrm{F,e}c}{m_\mathrm{e}c^2}) \right\},\\
	\label{Eq: p_F rho relation}
	\rho_\mathrm{e} &= \frac{64 \mu_\mathrm{e} m_\mathrm{p}p_\mathrm{F,e}^3}{\xi^3 h^3},
\end{align}
where $\mu_\mathrm{e}$ is the mean molecular weight per electron and $E_\mathrm{F,e}$ is the Fermi energy of electrons, which is related to $p_\mathrm{F,e}$ as
\begin{align}
	E_\mathrm{F,e}^2 = p_\mathrm{F,e}^2c^2 + m_\mathrm{e}^2c^4\left(1 + 2\nu \theta_\mathrm{D}\right).
\end{align}

\begin{figure}[htpb]
	\centering
	\includegraphics[scale=0.5]{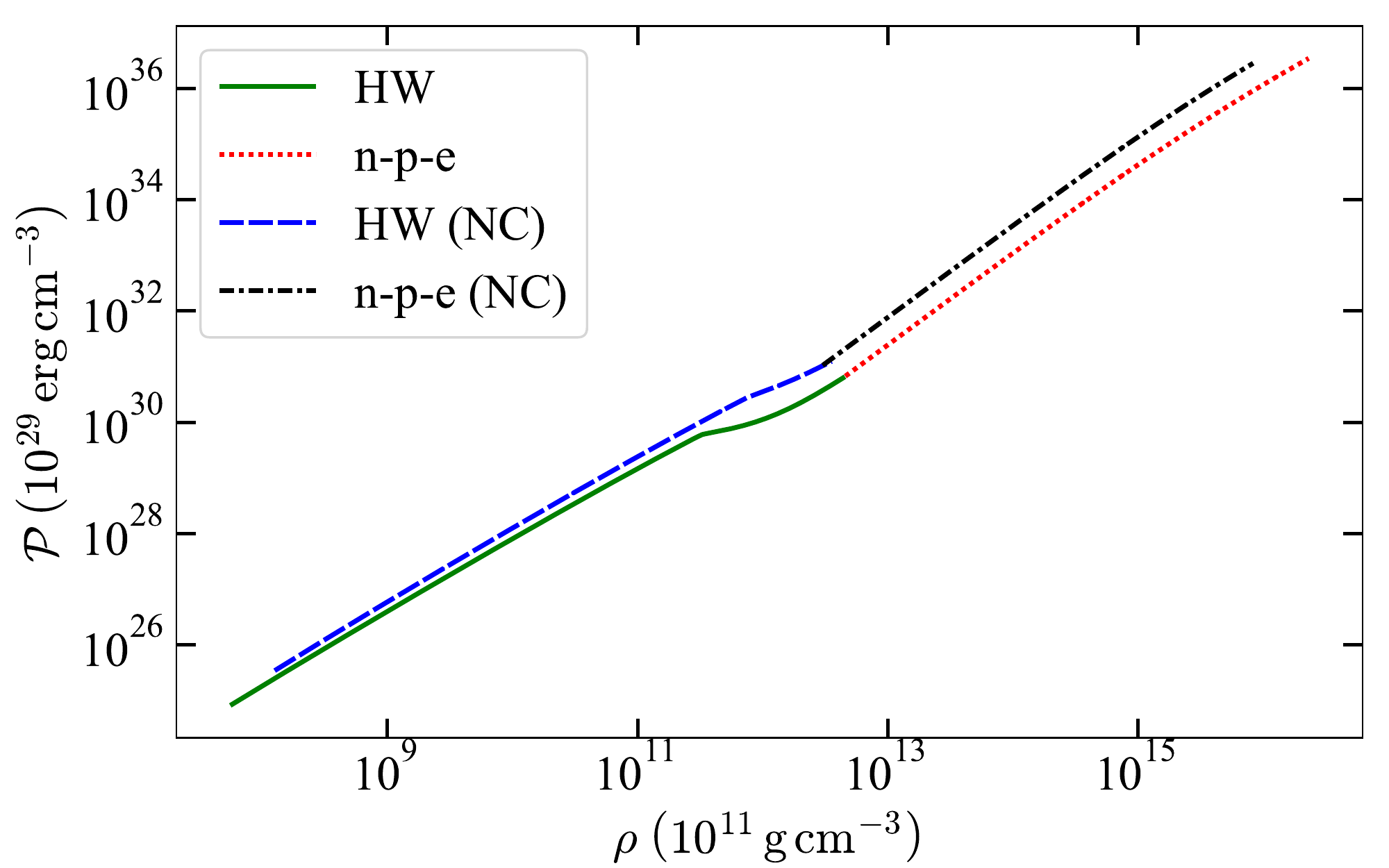}
	\caption{HW and n-p-e EoSs in the commutative and noncommutative formalisms.}
	\label{Fig: EoS_HW_npe_NC}
\end{figure}

Since, for the present purpose, we require the modified HW and n-p-e EoSs in the presence of NC, we also assume a similar form of pressure--density relation except that the various properties of the electron are now replaced by the same for the corresponding particle. After doing some simplifications using Equations~\eqref{Eq: theta_D n_e relation} and~\eqref{Eq: p_F rho relation}, we obtain
\begin{equation}\label{Eq: theta_D x_F relation}
\theta_\mathrm{D} = \frac{16}{\xi^3}\frac{x_\mathrm{F}^2}{\pi}.
\end{equation}
Note that we do not put any subscript for the electron in this equation, which means that it is valid for electrons, protons, and neutrons. We further denote the NC parameters of neutron, proton, and electron as $\theta_\mathrm{D,n}$, $\theta_\mathrm{D,p}$, and $\theta_\mathrm{D,e}$ respectively. Thus the modified HW and n-p-e EoSs are given by the same expressions of Equations~\eqref{Eq: HW} and~\eqref{Eq: npe}, except the pressures and energy densities of the respective particles are modified as follows:
\begin{align}
	\mathcal{P}_\mathrm{e} &= \frac{m_\mathrm{e}c^2\theta_\mathrm{D,e}}{2\pi^2\lambda_\mathrm{e}^3}\eta(x_\mathrm{F,e}), \quad \mathcal{P}_\mathrm{n} = \frac{m_\mathrm{n}c^2\theta_\mathrm{D,n}}{2\pi^2\lambda_\mathrm{n}^3}\eta(x_\mathrm{F,n}),\quad \mathcal{P}_\mathrm{p} = \frac{m_\mathrm{p}c^2\theta_\mathrm{D,p}}{2\pi^2\lambda_\mathrm{p}^3}\eta(x_\mathrm{F,p}),\\
	\epsilon_\mathrm{e} &= \frac{m_\mathrm{e}c^2\theta_\mathrm{D,e}}{2\pi^2\lambda_\mathrm{e}^3}\psi(x_\mathrm{F,e}), \quad \epsilon_\mathrm{n} = \frac{m_\mathrm{n}c^2\theta_\mathrm{D,n}}{2\pi^2\lambda_\mathrm{n}^3}\psi(x_\mathrm{F,n}), \quad \epsilon_\mathrm{p} = \frac{m_\mathrm{p}c^2\theta_\mathrm{D,p}}{2\pi^2\lambda_\mathrm{p}^3}\psi(x_\mathrm{F,p}),
\end{align}
where
\begin{align*}
    \eta(x_\mathrm{F}) &= \frac{1}{2}x_\mathrm{F}\sqrt{1+x_\mathrm{F}^2} - \frac{1}{2}\ln(x_\mathrm{F}+\sqrt{1+x_\mathrm{F}^2}),\\
    \psi(x_\mathrm{F}) &= \frac{1}{2}x_\mathrm{F}\sqrt{1+x_\mathrm{F}^2} + \frac{1}{2}\ln(x_\mathrm{F}+\sqrt{1+x_\mathrm{F}^2}).
\end{align*}
We already showed that if all the electrons reside only in the ground energy level, we require $\xi_\mathrm{e}\approx1.5$ to match the noncommutative EoS with the Chandrasekhar EoS at the low density~\cite{2021IJMPD..3050034K}. However, the corresponding parameters for neutron and proton ($\xi_\mathrm{n}$ and $\xi_\mathrm{p}$) remain arbitrary. We choose $\xi_\mathrm{n}$ and $\xi_\mathrm{p}$ in such a way that the maximum mass of NS in the mass--radius curve is above $2\rm\,M_\odot$, which we discuss in the next section. Thereby we calculate both the noncommutative HW and n-p-e EoSs when all the particles are in their respective ground levels (see Figure~\ref{Fig: EoS_HW_npe_NC}). Note that, the neutron drip density changes in the presence of NC, which was also shown earlier in the presence of strong magnetic fields forming Landau levels~\cite{2014PhRvC..89f5804V}.

\section{Mass--radius relation of noncommutativity inspired white dwarfs and neutron stars}\label{Sec: 4}
\begin{figure}[htpb]
	\centering
	\includegraphics[scale=0.45]{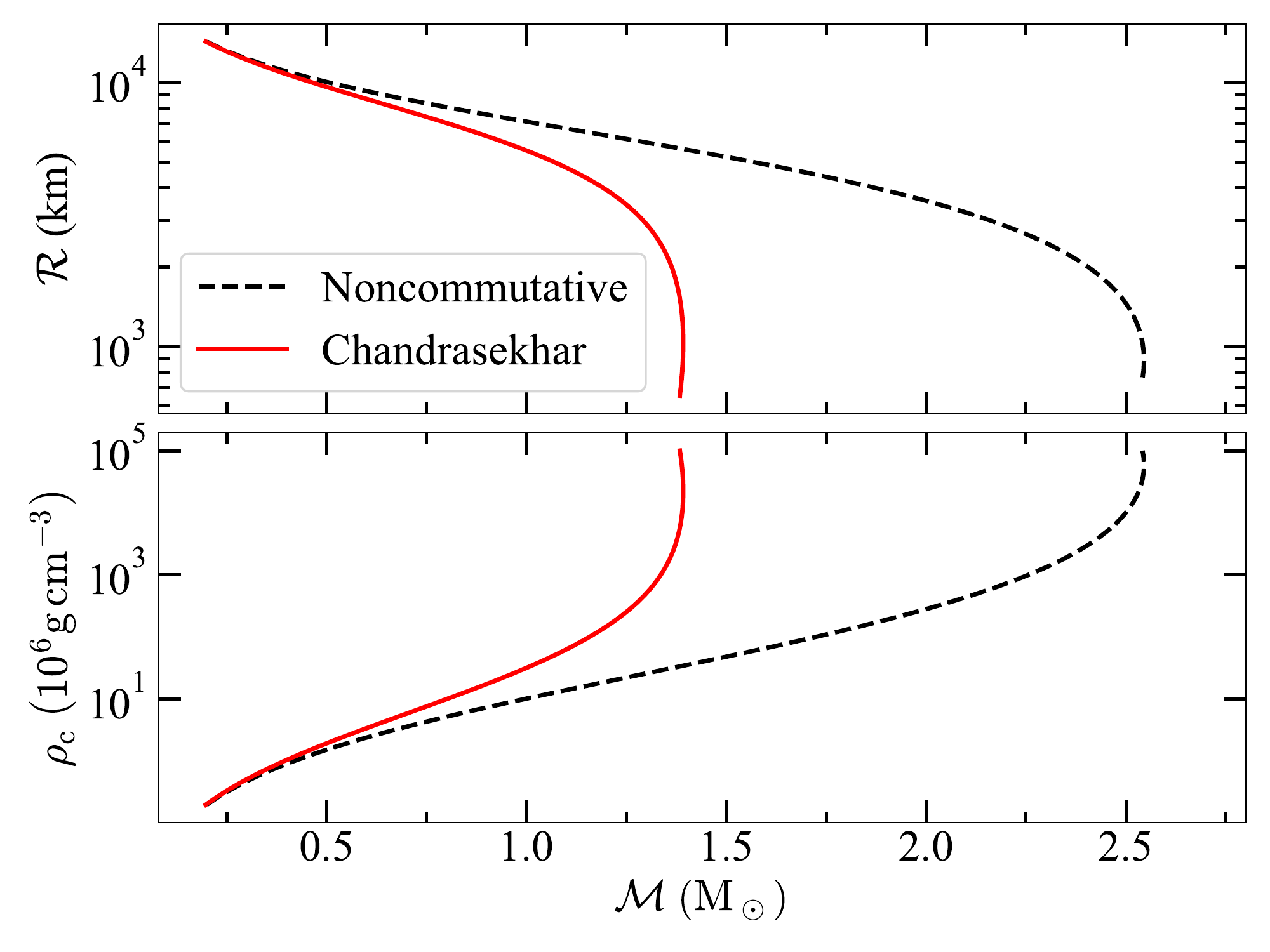}
	\caption{Upper figure: Mass--radius relation; Lower figure: variation of central density with the mass of WDs. Here $\mathcal{M}$ and $\mathcal{R}$ are the mass and radius of the star respectively.}
	\label{Fig: Fuzzy WD}
\end{figure}

We assume a semi-classical approach to obtain the mass--radius relations for WDs and NSs. In other words, we use classical pressure balance and mass estimate equations (also known as the Tolman–Oppenheimer–Volkoff or TOV equations) while the EoS is governed by the NC. The TOV equations are given by~\cite{2009igr..book.....R}
\begin{equation}\label{Eq: TOV}
\begin{aligned}
\dv{\mathtt{M}}{r} &= 4\pi r^2\rho,\\
\dv{\mathcal{P}}{r} &= -\frac{G}{r^2}\left(\rho+\frac{\mathcal{P}}{c^2} \right)\left(\mathtt{M}+\frac{4\pi r^3 \mathcal{P}}{c^2}\right) \left(1-\frac{2G\mathtt{M}}{c^2r}\right)^{-1},
\end{aligned}
\end{equation}
where $\mathtt{M}$ is the mass of the star inside a volume of radius $r$ and $G$ is the Newton gravitational constant. We earlier showed that NC is prominent when the inter-particle separation is less than the Compton wavelength of the respective particles~\cite{2021IJMPD..3050034K,2021IJMPD..3050101K}. When we consider the hydrostatic balance equations for the entire star having a macroscopic size, the length-scale of the stellar fluid is much larger than the Compton wavelength of the constituent particles. Thus, the TOV equations remain commutative in the semi-classical limit. Furthermore, when all the electrons reside in the ground energy level, we already found the mass--radius curve earlier~\cite{2021IJMPD..3050034K,2021IJMPD..3050101K}, and for recapitulation, we display it again in Figure~\ref{Fig: Fuzzy WD}. It is evident that NC inspired WDs can possess more mass than the conventional WDs following the Heisenberg algebra. The maximum mass of such a non-rotating WD is estimated to be around $2.6\rm\,M_\odot$, explaining the origins of many over-luminous SNe\,Ia.

\begin{figure}[htpb]
	\centering
	\includegraphics[scale=0.5]{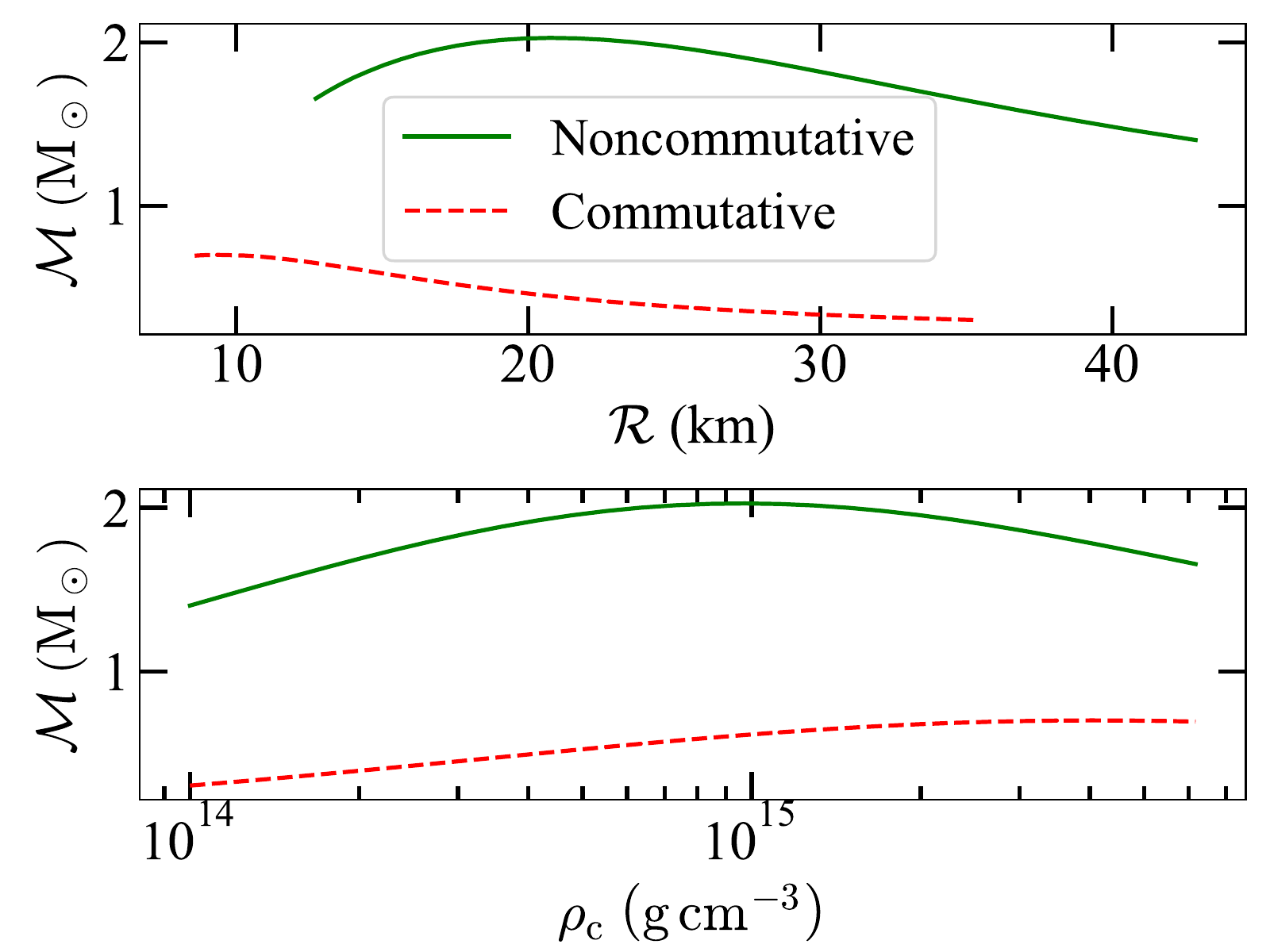}
	\caption{Upper figure: Mass--radius relation; Lower figure: variation of central density with the mass of NSs.}
	\label{Fig: MR_HW_npe_NC}
\end{figure}
\begin{figure}[htpb]
	\centering
	\includegraphics[scale=0.5]{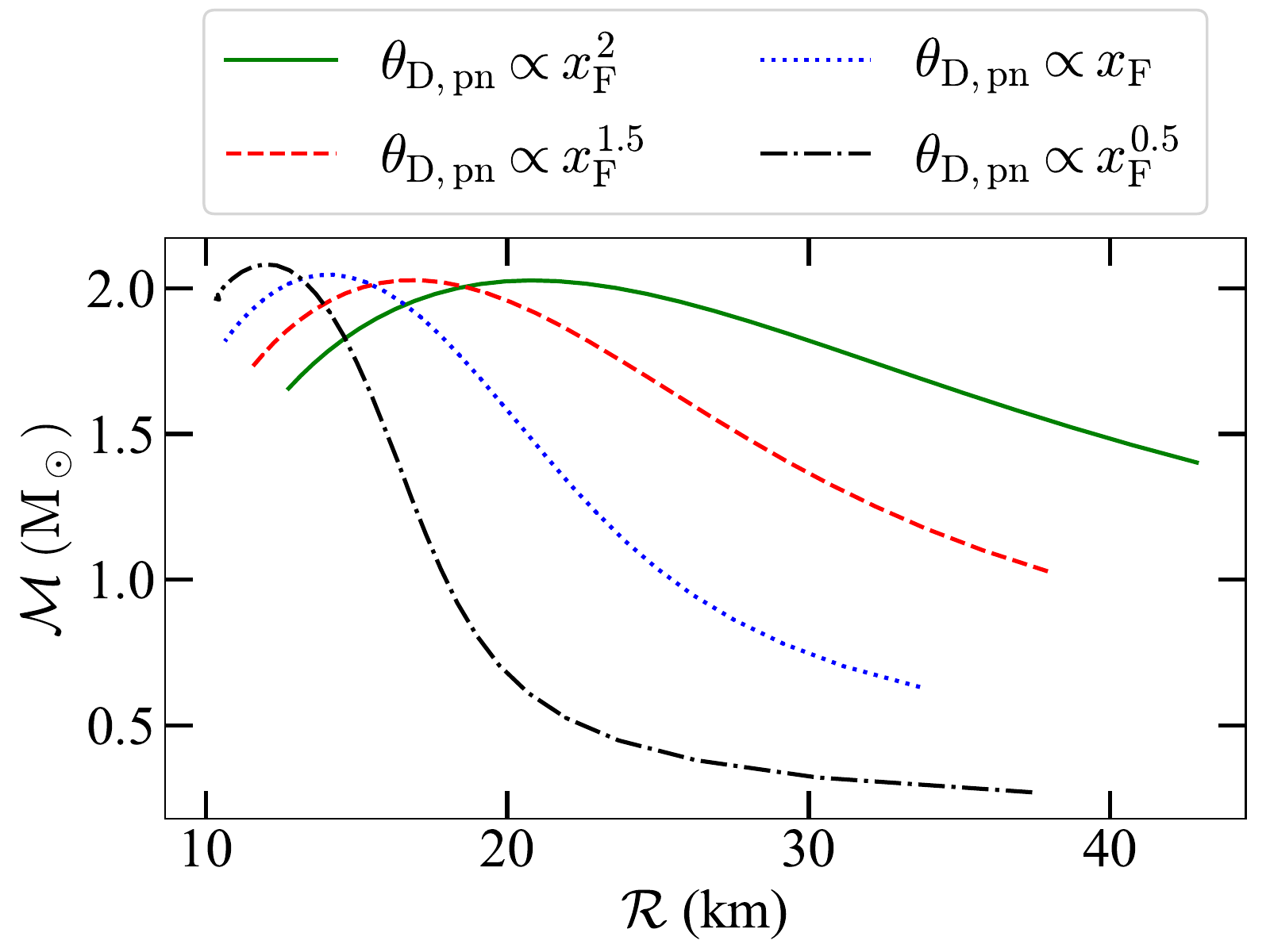}
	\caption{Mass--radius relations of NC induced NSs for various $\theta_\mathrm{D}-x_\mathrm{F}$ relations. $\theta_\mathrm{D,pn}$ means $\theta_\mathrm{D}$ for proton and neutron.}
	\label{Fig: MR_HW_npe_NC_all}
\end{figure}

In the case of a NS, $\rho_\mathrm{c}$ is high, and we employ a combination of HW and n-p-e EoSs to derive its mass–-radius relation, as illustrated in Figure~\ref{Fig: MR_HW_npe_NC}. In the commutative picture, the maximum mass turns out to be just $0.7\rm\,M_\odot$, while it is increased to about $2\rm\,M_\odot$ in the case of NC, which is supported by the observations of massive pulsars. However, the radius increases to $20\rm\,km$ in this situation, which is almost ruled out by existing GW observations~\cite{2018PhRvL.120q2703A,2020NatAs...4..625C,2022PhRvX..12a1058A}. Note that, the relation $\theta_\mathrm{D}\propto x_\mathrm{F}^2$ in Equation~\eqref{Eq: theta_D x_F relation}, is valid for electrons and we extrapolate it for neutrons and protons too. If we choose a different dependency of $\theta_\mathrm{D}$ on $x_\mathrm{F}$, the EoS alters and so as the mass--radius relation for NS. Figure~\ref{Fig: MR_HW_npe_NC_all} depicts several mass--radius relations for various powers of $x_\mathrm{F}$. It is evident that as the power decreases, the radius for maximum mass falls as well and when $\theta_\mathrm{D}\propto \sqrt{x_\mathrm{F}}$, the maximum mass is about $2.08\rm\,M_\odot$ with radius being $12\rm\,km$. These masses and radii obey the observational bounds of NSs, and hence such an EoS is a realistic one.

\section{Conclusions}\label{Sec: 5}

For a long time, scientists have been fascinated by the possibility of massive WDs and NSs from several direct or indirect observations. Various ideas, such as magnetic fields and rotation, modified gravity, etc., have been thoroughly investigated in recent years. Rotation can explain massive NSs, which, however, alone fails to elucidate the massive WDs with a mass of about $2.8\rm\,M_\odot$. High magnetic fields can, in principle, explain both these massive objects. Nonetheless, the maximum field that a compact object can possess is always a source of contention. Similarly, despite the fact that modified gravity can explain such high masses, it has so far been impossible to identify the most appropriate one from the hundreds of such modified gravity models. In this regard, each of these theories suffers its own limitations.

In the context of astronomical objects, the concept of NC is relatively new. With the exception of a few applications on black holes and wormholes, it has received little attention in astrophysics. We earlier self consistently used NC for the first time to explain the super-Chandrasekhar WDs~\cite{2021IJMPD..3050034K,2021IJMPD..3050101K}. We first employed a basic planar NC model and later used a squashed fuzzy sphere model to modify the EoS of the degenerate electrons present in a WD. This modification leads to increasing the mass of a WD. If the electrons solely occupy the ground energy level, i.e. NC is the strongest, the new mass-limit of WD turns out to be about $2.6\rm\,M_\odot$. As NC weakens and electrons occupy higher energy levels, this mass-limit decreases. It is to be noted that the effect of NC is only prominent at sufficiently high densities and negligible at low densities. Hence, our model supports the observed bigger WDs, which generally have very low densities, and it does not violate any observable at such low densities. We have already established that the strength of NC depends on the length scale of the system. If the inter-particle separation distance is smaller than the Compton wavelength of the corresponding particle, NC starts becoming prominent~\cite{2021IJMPD..3050034K}. Furthermore, NC does not have any classical effect, unlike magnetic fields (i.e. field pressure, tension, etc.), and hence the problem of instabilities that occurred in magnetic fields does not arise in the case of NC, making the NC model preferable over that of magnetic fields.

In this article, we have extrapolated NC to higher densities and investigated for its effect on the structure of NSs. For simplicity, we have only considered the effects of neutrons, protons, and electrons and assumed they are non-interacting. In commutative physics, it is well known that such an EoS gives the maximum mass of a NS to be about $0.7\rm\,M_\odot$~\cite{compact}. However, current observations demand the maximum mass of a non-rotating NS has to be at least $2\rm\,M_\odot$~\cite{2018ApJ...852L..25R,2019PhRvC..99e2801M,2020MNRAS.499L..82M}. Once we introduce NC, we have found that even such non-interacting particles can constitute an EoS which generates NS with a mass of about $2.1\rm\,M_\odot$ and radius $12\rm\,km$. Such an EoS is perfectly legitimate with the current observation constraints. Note that we have only considered the case where all the particles are in their respective ground energy levels, which is the scenario for the strongest NC. One can, in principle, consider higher occupancy in the energy levels. However, it only reduces the mass of the NS, as we have seen in the case of WDs~\cite{2021IJMPD..3050101K}, and the maximum mass could fall below $2\rm\,M_\odot$ and those cases would be unrealistic. In such instances, one must account for the interactions that may occur between the various particles at these high densities; which however is beyond the scope of this paper. In such cases, even the EoSs, which are considered non-physical, might not be ruled out if they are affected by NC. In the future, GW observations may detect numerous massive WDs and NSs~\cite{2019MNRAS.490.2692K,2020ApJ...896...69K,2021MNRAS.508..842K}, allowing us to constrain more EoSs and examine the NC effect on these compact objects more closely. If observed masses and radii of WDs and NSs follow the respective predicted mass--radius relations based on NC, it would be a direct confirmation for the existence of NC at scales far away from the Planck scale.

\acknowledgments{SK would like to acknowledge support from the South African Research Chairs Initiative of the Department of Science and Technology and the National Research Foundation.}

\begin{adjustwidth}{-\extralength}{0cm}

\reftitle{References}


\bibliography{bibliography}

%


\end{adjustwidth}
\end{document}